\documentstyle[referee]{laa}
\thesaurus{08.19.1}
\title{The fluctuating gravitational field in inhomogeneous and clustered 
self-gravitating systems.\\ } 
\author{A. ~Del Popolo\inst{}} 
\offprints{A. ~Del Popolo}
\institute{Istituto di Astronomia dell'Universit\`a di Catania, \\
Citt\`a Universitaria, Viale A.Doria, 6 - I 95125 Catania, Italy}
\date{}
\begin{document}
\maketitle
\begin{abstract}
In this paper I extend the results of Ahmad \& Cohen (1973), regarding 
the study of the probability distribution of the stochastic 
force in homogeneous gravitational systems, to  
inhomogeneous gravitational ones.
To this aim, I study the stochastic force 
distribution using N-body realizations of Plummer's spherically symmetric 
models. 
I find that the stochastic force distribution obtained for 
the evolved system is in good agreement with 
Kandrup's (1980) theory of stochastic force in inhomogeneous systems. 
Correlation effects 
that arise during the evolution of the system of particles 
are well described by Antonuccio-Delogu \& Atrio-Barandela's (1992) 
theory.    
\end{abstract}
\begin{flushleft}

{\bf 1. Introduction \\}

\end{flushleft}

\vspace*{0.5 mm}

\vspace*{0.5 mm}
The analysis of the dynamics of stellar systems 
such as globular clusters or clusters of galaxies has 
shown that the gravitational stochastic force plays a fundamental 
role in their evolution. 
In these systems the 
stochastic force, arising from statistical fluctuations in the number of 
neighbours of a test star, perturbs the stars orbits from 
the orbits they would have if the 
density distribution in the system were perfectly smooth. The first consequence 
produced by the stochastic force  
is the existence of a frictional force that implies a preferential deceleration 
of a particle in the direction of motion (Chandrasekhar 1943, Ahmad \& 
Cohen 1974). 
The existence of the 
stochastic force is due to the discreteness of gravitational systems, i.e. 
to the fact that the   
mass is concentrated into discrete objects 
like stars. The gravitational field inside such systems 
can be in principle 
computed by summing up the potentials of all 
stars. Obviously, this is not practicable when the number 
of stars constituting an object (e.g a globular cluster or a galaxy) is large 
and so it is 
generally assumed 
that the potential can be expressed as an uncorrelated sum of two fields: 
a mean field obtained from the smoothed mass distribution and a 
rapidly varying random fluctuation, representing stochastic deviations from 
the mean field conditions. \\
The mean field can be determined through the Poisson's equation 
when the density of the system is given, while the determination of the 
fluctuating field is a more complex task. \\
The analysis of fluctuations has been formulated through different 
approaches. One of them is the 
two body approximation (see Kandrup 1980). In this approach, 
fluctuations are approximated 
by a sum of binary interactions of the test particle 
with the remaining particles. As a consequence, fluctuations 
are required to have a timescale, $T$, small in comparison with  
any other relevant timescale, (the dynamical timescale and the 
relaxation timescale). The main assumptions of this approach
are that the probability distribution verifies a 
Fokker-Planck equation and that the fluctuations 
in velocity of a 
test star are due to the sum of binary encounters with neighbouring 
stars. Finally, there is the inherent assumption that the stochastic 
force on a test star, at a given time, is equal to the sum of 
the effects due 
to each field star. As a result, in the limit of a small 
angle of scattering there is a logarithmic divergence in the velocity 
fluctuation  
$ \frac{ \delta v^{i} }{ \delta t}$. Physically, this arises because the sum 
of the forces acting on a test star in an infinite system 
diverges. The divergence can be solved introducing a cut-off at  
a distance of the order of the mean particle distance, $ r_{med}$ 
(Kandrup 1980). \\
To solve these problems it is necessary to use 
a full stochastic treatment which 
takes correctly into account the effects 
of the field stars whose distance from a 
test star is greater than the interparticle 
separation. \\
The stochastic treatment of gravitational fluctuations is 
connected with the evaluation of a density probability, $ W( {\bf F})$, 
i.e. the probability that a test star is subject at a given time 
to a stochastic force, $ {\bf F}$. In the simplest case (homogeneous, 
unclustered, infinite system) the distribution $ W({\bf F})$ was 
calculated by Chandrasekhar \& von Neumann (1942) (hereafter CN), while   
the theory of stochastic force in inhomogeneous, unclustered, infinite   
systems was developed by 
Kandrup (1980). 
The results of Kandrup (1980) showed that the 
CN theory is in reality independent of 
the condition of uniform density. 
Also in inhomogeneous systems, like in homogeneous, the 
cancellation of contributions to the force from distant field stars is shown 
to yield essentially negligible contribution to the total magnitude 
of the   
fluctuating force (Kandrup 1980) (while the field stars 
have a nontrivial role in the determination of the 
distribution of stochastic force in the weak force limit)
and so the basic results given by 
CN are not dependent 
on the density profile and can be easily extended to inhomogeneous systems. \\
A comparison of the two-body approximation (with cut off at radius 
$ r_{med}$) with the full stochastic theory shows that for large forces 
the two treatments are equivalent while, in the limit of weak 
forces, they disagree 
because distant field stars play a nontrivial role in the origin 
of the stochastic force. \\
A self-consistent mean field theory, like   
CN and 
Kandrup's theory,  
can describe correctly the 
stochastic force in a system only if  
there are no correlations among the positions of particles 
so that 
the probability 
of finding one particle at a position $ {\bf x}$ and another at $ {\bf y}$ is 
given by: 
\begin{equation}
n_{2} ({\bf x}, {\bf y}) = n_{1} ({\bf x}) n_{2} ({\bf y})  
\end{equation}
where $ n_{1} ({\bf x}) $ is the probability density of finding 
a particle at $ {\bf x}$. This 
assumption requires that the potential experienced by a test 
particle is on average larger than the binary potential 
$ \frac{G m}{R}$ ($m$ is the particle mass and $R$ the radius 
of the system) connected with binary encounters. 
Only with the simplifying hypothesis of weakly clustered 
systems it is possible to extrapolate the results of Kandrup (1980) to 
clustered systems (Antonuccio-Delogu \& Atrio-Barandela 1992, hereafter 
AB92).  \\
Numerical experiments on the CN distribution were done  
by Ahmad \& Cohen (1973). As previously stressed, this distribution 
is a theoretical description of the stochastic force in the homogeneous systems 
only. In this particular case they found a good agreement of numerical 
experiments 
with the theoretical distribution. \\
The only test of the theoretical distribution of stochastic force in 
inhomogeneous systems (Kandrup 1980) and clustered systems 
(Antonuccio-Delogu \& Atrio-Barandela 1992)  
is that performed  
using unevolved systems by A. Del Popolo (1995), who found a good  
agreement of the theoretical distribution with that derived 
from numerical 
experiments. 
In the quoted paper the experimental stochastic force was calculated in  
unevolved inhomogeneous and clustered systems of particles and 
directly compared with Kandrup's and AB92 distributions. 
Something similar
was performed by Hunger et al (1965). In their paper numerical experiments 
were performed to verify 
the theoretical stochastic distribution of electric field magnitude 
into a homogeneous system of particles (Holtsmark distribution) without 
evolving the N-body system.\\ 
This procedure can be justified as follows:\\
the force per unit mass acting on a test star is given 
by:
\begin{equation}
{\bf F} = -G \sum_{i=1}^{3} \frac{m_{i}}{|{\bf r}_{i} -{\bf r}|^{3}}
({\bf r}_{i} -{\bf r}) 
\end{equation}
where the sum is extended to the N stars of the system, 
$ m_{i}$ is the mass of the i-th field star and $ r_{i}$ is its distance 
relative to the origin. Owing to the motions of the stars, the force acting 
on a point P will change with time. If $ \psi( {\bf F}_{0}; {\bf F}, t)$ 
denotes the probability that a force of intensity 
$ {\bf F}$ acts at P after 
a time $ t$ then: 
\begin{equation}
\psi( {\bf F}_{0}; {\bf F}, t) \rightarrow W(F) \hspace*{1.0cm} t>>T
\end{equation}
(Chandrasekhar \& von Neumann 1942), 
where $ {\bf F}_{0}$ is the force at $t=0$,
\begin{equation} 
T \simeq (\frac{2 \pi G m}{3 < v^{2}>)})^{1/2}  
\frac{|{\bf F}|}{[Q_{H}^{3/2} + |{\bf F}|^{3/2}]}
\end{equation}
is the mean life of the state 
$ {\bf F}$, $ Q_{H} = 2.6031 G m n^{2/3} $, $n $ is the number density and 
$ \sqrt{< v^{2}>}$ is the root mean squared velocity of the field particles.
By a state $ {\bf F}$ I mean (according to Chandrasekhar \& von Neumann 1942) 
that at a fixed point P
a force per unit mass of intensity $ {\bf F}$ is acting at an instant $ t$. 
After a sufficient length of time, the force acting 
on P will be uncorrelated with that acting at time $ t=0$. 
If the process that generates the change in time of the 
force is a Markoff process 
the correlations 
between the force $ {\bf F} (t_{1})$ and $ {\bf F} (t_{2})$ 
acting on the same point, but at two different instants, decrease exponentially 
as $ \exp{-\frac{t}{T}}$. In this way the notion of the mean life of the 
state $ {\bf F}$ is made clear. 
Then, for a time greater than the mean life of 
the state $ {\bf F}$ the stochastic force distribution reduces to a stationary 
distribution (Chandrasekhar \& von Neumann 1942) $ W(F)$ 
that is a function of the density $ n$ of the system. \\ 
If, for example, the particles are distributed with a probability density 
in the configuration space given by: 
\begin{equation}
\tau(r) = \frac{a}{r^{p}}
\end{equation}  
where $ a$ and $ p$ are two constants, 
the theoretical distribution of the stochastic force 
is: 
\begin{equation}
W(F) = \frac{2 F}{ \pi} \int_{0}^{\infty} t dt \sin( t F) 
\exp \left[-\frac{\alpha}{2} ( G mt)^{(3-p)/2} 
\int_{0}^{\infty} \frac{ dz (z -\sin z )}{z^{(7-p)/2}}\right]
\end{equation}
where $ \alpha$ is connected to the number density $ n$ 
($ \alpha = 4 \pi n$ for $ p=0$). As shown in 
Sect.~2 of this paper, in the case of a Plummer model $ W(F)$ is 
again function of the density (Eq.~10). If the system is generated using a 
distribution function that satisfies Vlasov equation 
(as in the case of Plummmer's
model) it remains time-stationary on time interval 
$ t_{cross}<\Delta t<< t_{relax}$, where $ t_{cross}$ is the crossing time  
and $ t_{relax}$ the relaxation time. In the case of a Plummer model 
this is shown in a paper by 
L.~Hernquist (1987). This paper shows how the density profile 
remains unchanged during time evolution over 50 crossing times. 
Then the shape of $ W(F)$ remains unchanged 
in the quoted interval until correlations effects are developed. This 
justifies the procedure by A. Del Popolo's (1995) paper. \\
The aim of A. Del Popolo's (1995) 
paper was to show that, given an inhomogeneous density profile, 
the stochastic force can be described by Kandrup's or the AB92 theory.  
Different is the purpose of this paper. 
Here I study the stochastic force in an evolving inhomogeneous system
to find if and when Kandrup's and AB92 theories describe 
correctly the stochastic force in the system. This paper deals 
with the stochastic force in real systems. It is then necessary 
to distribute the system particles in 
an inhomogeneous  
initial configuration and to evolve it.  
The evolution of the system 
is necessary at least for two reasons: \\
firstly, the structure of real systems like globular clusters 
is fundamentally due to their 
dynamical evolution.
To compare the theoretical 
results to real inhomogeneous 
systems it is then necessary to evolve an inhomogeneous initial configuration 
and to calculate the stochastic force during the evolution;  \\
secondly, the evolution of a system produces correlations in the 
configuration space. Self-gravitating systems are unstable with respect 
to the development of macroscopic correlations.  
Correlations produce an enhancement in the probability that 
a test star experiences large forces (AB92) and this 
produces a change in the stochastic force distribution 
$ W(F)$. \\
The aim of this paper is to study the stochastic force 
in real, evolved systems. \\
In 
this paper it is  shown 
that the stochastic force in inhomogeneous evolved systems is 
well described by Kandrup's theory as long as correlations are 
developed. When correlations are present a better 
description of the stochastic force can be obtained using 
AB92 theory.  
This was done by numerically evolving an isotropic 
Plummer model containing 8000 
particles for 150 dynamical times. The stochastic force distribution 
was calculated during 
the evolution on a test star set at the system centre and it was 
compared with Kandrup's theoretical distribution. The force was also 
calculated using 1000 points distributed into a sphere of radius $ 0.01R$ 
(R is the maximum radius)
centred on the system centre. The disagreement observed on occasion 
between Kandrup's distribution 
and the experimental one was removed using Antonuccio-Delogu \& 
Atrio-Barandela theory.\\
The paper has the following structure: \\
in Sect.~2 I describe how the system was built and how 
the stochastic force was calculated. In Sect.~3 I show the results of the  
numerical experiments when correlations are not present. 
In Sect.~4 the disagreement between Kandrup's distribution 
and the experimental one, observed in the presence of correlations, is removed 
using the AB92 theoretical distribution.\\      
\begin{flushleft}
{\bf 2. Stochastic force in inhomogeneous systems\\} 
\end{flushleft} 
To calculate the stochastic force in an inhomogeneous system 
I used an initial configuration in which particles were 
distributed according to a  
Plummer model with density profile:
\begin{equation}
\rho(r) = \frac{3 M}{4 \pi r_{0}^{3}} 
\frac{1}{\left[1+(r/r_{0})^{2}\right]^{5/2}} \label{eq:plum}
\end{equation}
where $ M$ is the total mass of the system and $ r_{0}$ is the scale length.
The isotropic distribution that reproduces the mass distribution 
given by Eq.~\ref{eq:plum} is: 
\begin{equation}
f(E)= \frac{\sqrt{2}}{378 \pi^{3} G r_{0}^{2} \sigma_{0}} 
(-E/\sigma_{0}^{2})^{7/2} 
\hspace{1.0cm} -6 \sigma_{0}^{2}\leq E \leq 0 \label{eq:dist}
\end{equation}
(Eddington 1916) \\
where $ \sigma_{0}$ is the central velocity dispersion and $ E$ 
is the total energy.  
This distribution function was chosen for three reasons: \\
a) being it a function of an integral of motion according 
to Jeans theorem it must be a steady state 
solution to Vlasov equation and in absence of numerical 
errors and dynamical instabilities (Merrit \& Aguilar 1985)
it remains stationary; \\
b) according to Doremus-Feix-Baumann theorem (1971) and to Antonov's 
second law the Plummer model is stable. \\
On the other hand Antonov (1973) 
concluded that any spherical system composed entirely of stars 
on radial orbits is unstable. This conclusion can be extended to  
anisotropic spherical systems in which, as shown by   
Henon's (1973 b) and Barnes' (1986) numerical 
simulations, there exists 
a so called radial-orbit-instability that leads to a triaxial or 
bar-like final 
configuration. In the case of Plummer's isotropic model 
(Eq.~\ref{eq:dist})  
this problem is not 
present in my simulations, while an anisotropic Plummer model 
(Merrit 1985)
shows the radial orbit instability;   \\
c) the density distribution in a Plummer model is similar 
to the density distribution 
of many real systems. Plummer (1911) showed that the density 
distribution given by Eq.~\ref{eq:plum} provides a good fit to 
the density profile of  
globular clusters. \\
\begin{figure}
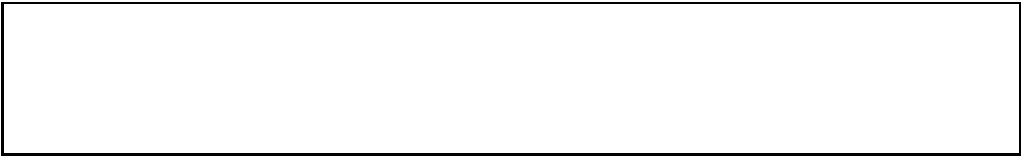

\picplace{2.0cm}
\caption[]{Comparison of the theoretical stochastic force distribution 
(solid line) 
with the experimental distribution (histogram) for an inhomogeneous 
system of 8000 particles evolved over 25 dynamical times. 
The force is calculated 
using a test particle set at the centre of the system.}
\end{figure} 
The initial conditions were generated 
from the distribution function given in Eq.~\ref{eq:dist}
assuming a cut-off radius $ R=1$, the mass of the system $ M=1$, 
$ r_{0} =0.15$ and $ G=1$. All the particles had  
equal mass. 
\begin{figure}
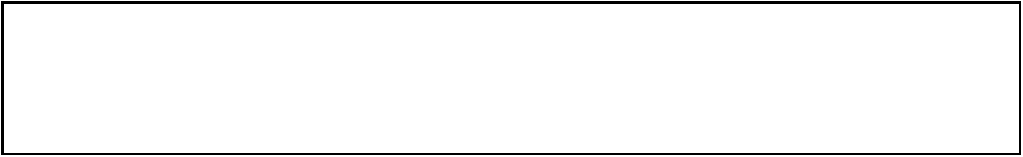

\picplace{2.0cm}
\caption[]{Same as Fig.~1 but now the system is evolved over 50 
dynamical times.}
\end{figure} 
To have a system whose total mass is contained into a unitary 
sphere, Eq.~\ref{eq:plum} was renormalized and 
consequently also the potential of the system 
which is obtained from Eq.~\ref{eq:plum} through Poisson's equation. 
The system of 8000 particles 
was evolved over 150 dynamical times using a N-body code 
(L.Hernquist 1987). 
During 
the evolution of the system the total force acting on a test point at the 
centre of the system was sampled 
every $ \frac{1}{20}$ of a dynamical time. 
The stochastic force, $ {\bf F}_{stoch}$, was calculated observing 
that at the centre of a spherical system we have: 
\begin{equation}
{\bf F}_{tot} = {\bf F}_{stoch}+{\bf F}_{med} = {\bf F}_{stoch}
\end{equation}
because the mean field force, $ {\bf F}_{med}$, is equal to zero.
The force was calculated on a point at the centre of the system 
because theoretically Kandrup's distribution gives the probability 
distribution of the stochastic force for a particle at the centre only. 
When points displaced away from 
the centre are used the stochastic force distribution 
must be calculated as follows: \\
a) the stochastic force should be calculated by subtracting the mean field 
force from the total force: 
\begin{equation}
{\bf F}_{stoch}={\bf F}_{tot}-{\bf F}_{med}
\end{equation}
b) the theoretical distribution must be numerically simulated as done by Ahmad 
\& Cohen (1973). \\
\begin{figure}
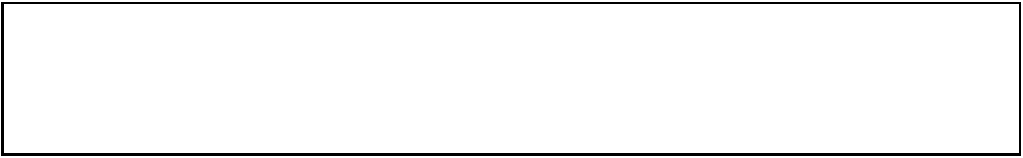

\picplace{2.0cm}
\caption[]{Comparison of the theoretical stochastic force (solid line) 
with the experimental distribution (histogram) for an inhomogeneous 
system of 8000 particles evolved for 20 dynamical times. The force 
is calculated as described on the 
text using 1000 test particles set into a sphere of radius $0.01 R$ 
centred in the centre of the system.}
\end{figure} 
The stochastic force was also calculated using a series of snapshots 
of the evolved system. In each case the force was computed  
using 1000 randomly distributed test points in a sphere of radius $0.01R$. 
The force on each 
test point was obtained by subtracting the mean 
field force from the total force.
In both cases the stochastic 
force was expressed in terms 
of $ \frac{G m}{r_{0}^{2}}$, where $m $ is the mass 
of one particle.  
\begin{figure}
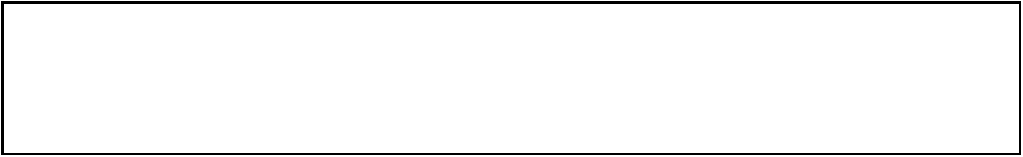

\picplace{2.0cm}
\caption[]{Same as Fig.~3 but now the system was evolved till over 80 dynamical 
times.}
\end{figure} 
The theoretical distribution for the density 
profile given in Eq.~\ref{eq:plum} was calculated 
following the same technique 
followed in the Appendix A of 
A. Del Popolo (1995): 
\begin{equation}
W (F) = A \cdot F \int_{0}^{\infty} \rho d\rho \sin (\rho F) 
\exp \left[-\frac{\alpha}{2} ( G m\rho)^{3/2} B(\rho)\right]
\end{equation}
where A is a constant of normalization, $ \alpha = \frac{3 N}{ R^{3} }
\left[1+(R/r_{0})^2\right]^{3/2}$, $ N$ is the number of particles 
and $ B(\rho)$ is given by:
\begin{equation}
B(\rho) = \int_{0}^{\infty} \frac{z-\sin z}{z^{7/2}} 
\frac{d z}{(1+\frac{Gm \rho}{r_{0}^2 z})^{3/2}}
\end{equation}
This theoretical distribution is slightly different 
from that given by Kandrup (1980) for a power-law density profile. 
In particular the parameter $ \alpha$ and 
$ B( \rho)$ are different from the previous ones. The $ \alpha$ parameter 
is strictly connected to the density in the system and on it depends 
the magnitude of the stochastic force per unit mass, $ <|F_{stoch}|>$. 
As shown by Kandrup, 
this has the magnitude which may be expected from the interaction 
of the test star with a few nearby field stars. In the case of the 
Plummer model this is an order of magnitude greater than the value obtained in 
the homogeneous case, for which is:
\begin{equation}
<|F_{stoch}|> = 8.879 G m n^{2/3} \simeq 0.4
\end{equation}
where $m$ is the mass of one particle, $n$ the mean density (the units 
previously introduced were used).
The theoretical distribution of the stochastic force was finally 
compared with the histogram of force 
obtained from the evolved system. \\
\begin{flushleft}
{\bf 3. Results of the numerical experiments\\}
\end{flushleft}
The results obtained from the numerical experiments performed 
are shown 
in Figs.~1-8.  
Fig.~1 and Fig.~2 show the distribution 
of the 
stochastic force calculated using a test point at the centre of the system. 
\begin{figure}
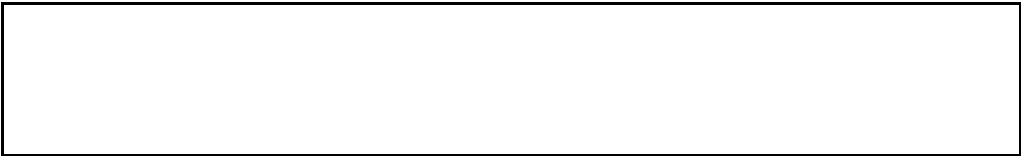

\picplace{2.0cm}
\caption[]{Comparison of the theoretical stochastic force (solid line) 
with the experimental distributions for an inhomogeneous 
system of 8000 particles evolved over 85 (solid line histogram), 
100 (dashed line histogram), 
150 (dotted line histogram) dynamical times. 
The force is calculated 
using a test particle set at the centre of the system. The disagreement 
between experimental and theoretical distributions is due to 
the correlations in configuration space that arise during the evolution 
of the system.}
\end{figure}
Fig.~1 is obtained from the system evolved over 25 dynamical times. 
It shows a good agreement of the theoretical distribution (solid line) 
with the experimental 
distribution of force obtained from the system (histogram). 
Fig.~2 is the same as 
the previous one but in this case the system is evolved over 50 dynamical 
times. Fig.~3 and 4, are obtained from the system of particles 
evolved respectively over 20 and 80 dynamical times. Now  
the force 
is calculated using 1000 points randomly distributed into a 
sphere of radius $0.01 R$. Also in these cases the theoretical distributions 
(solid line) are in good agreement with the experimental ones
(histogram). \\
\begin{figure}
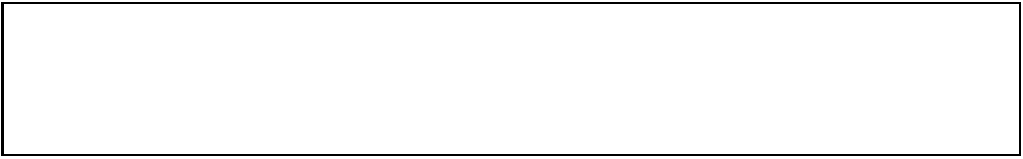

\picplace{2.0cm}
\caption[]{Distribution of cosines of the angle between pairs of particles 
as measured from centre of the system for a system evolved over 
100 dynamical times.}
\end{figure}
During the evolution of the system I observed  
a disagreement between 
the theoretical and the experimental distributions, plotted in Fig.~5.  
This figure shows the comparison between the theoretical 
stochastic force (solid line) and the experimental distributions  
in a system evolved over 85 (solid line histogram), 
100 (dashed line histogram), 150 (dotted line histogram) 
dynamical times and with the stochastic force 
calculated 
using a test point at the centre of the system. 
The disagreement is observed when the system is 
evolved over times larger than 80 dynamical times. 
Disagreements of this kind were found by Ahmad \& Cohen (1973) 
in their simulations of homogeneous systems and were attributed to correlations 
in the configuration space. Then it is interesting to test this hypothesis 
also in the inhomogeneous system I evolved.\\ 
~\\
\begin{flushleft}
{\bf 4. Correlations and stochastic force.\\}
\end{flushleft}
Correlation effects in gravitational system were studied in several papers 
(Prigogine \& Severne 1966; Gilbert 1970; 
AB92). Already Gilbert (1970) 
predicted that correlations produce an 
enhancement of the stochastic force, this conclusion being 
confirmed in 
the study of AB92. 
Thus a cause of the 
disagreement observed in my simulations 
between theoretical and experimental stochastic 
force could be the correlations arising during 
the evolution. To test this hypothesis and verify the possible  
presence of correlations, the test of the 
distribution of cosines of the angle between pairs of particles 
was used (Miller 1971). 
If in a system correlations are not present, the distribution of cosines 
is uniform between -1 and 1 with a probability density of 
0.5. A larger value 
for a large value of the cosines indicates the presence of clustering. 
The test for the system evolved over 85, 100, 150 dynamical times was 
performed. In all three cases there was evidence for clustering 
in the configuration space. The result of Miller's test for the 
system evolved over 100 dynamical times is   
shown in Fig.~6. For $ cos \theta \simeq 1$ the density distribution 
is larger than 0.5 indicating a correlation in the configuration space. 
To measure quantitatively the correlations in the system of particles it 
is necessary to measure the two points correlation function $ \xi$. 
This is possible using the counts in cell analysis 
(Ripley, 1980; Mo, 1991). 
Once known the two points correlation function the theory 
developed in AB92 was used to calculate the theoretical distribution 
of a correlated system. The equation used to calculate the 
distribution of force is:
\begin{eqnarray}
W_{N}(F) = 4 \pi^{2} |{\bf F}|^{2} W_{N} ({\bf F}) = 
\frac{2 F}{ \pi} \int_{0}^{\infty} t \sin( t F) 
\left[ \frac{\alpha}{N} (  G m t )^{ (3-p)/2} 
\int_{\frac{G m t }{ R^{2}}}^{\infty} \exp( -\frac{G m t}{ r_{0}^{2} z})
\frac{ \sin z }{z^{(7-p)/2}}\right]^{N} \nonumber \\
\cdot \left[1 + \frac{1}{2}( 1- \frac{1}{N}) \frac{\Sigma(t)}{ A_{2} (t)}
\right]
\end{eqnarray}
where $ \Sigma(t)$ and $ A_{2} (t)$ are given in the 
quoted paper (Eq.~34 and Eq.~32 respectively).  
\begin{figure}
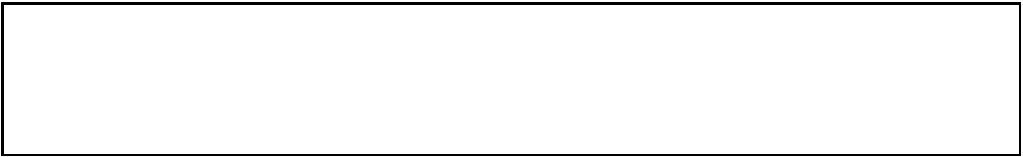

\picplace{2.0cm}
\caption[]{Comparison of the theoretical stochastic force (solid line) 
with the experimental distribution (histogram) for an inhomogeneous 
system of 8000 particles evolved over 85 dynamical times. 
The disagreement between the two distribution can be removed 
when correlation effects are taken into account 
and the theoretical distribution given by Kandrup is 
replaced by Antonuccio-Delogu \& Atrio-Barandela distribution 
for clustered systems (dotted line). }
\end{figure}
In Figs.~7 and 8   
this theoretical distribution (dotted line) is compared with 
the experimental ones 
(histogram) and with Kandrup's distribution (solid line) in the case    
of a system evolved over 85 and 100 dynamical times, respectively. 
\begin{figure}
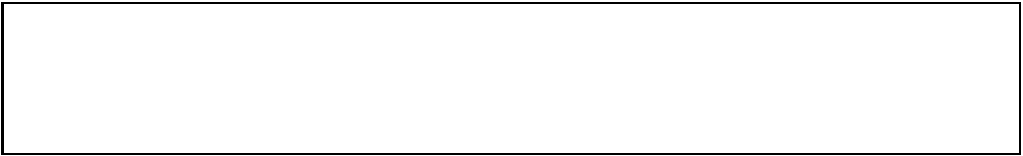

\picplace{2.0cm}
\caption[]{Same as Fig.~7 but now the system was evolved over 100 
dynamical times.}
\end{figure}
The disagreement between Kandrup's distribution 
and the experimental one is due, as previously written, 
to the correlations that arise during the evolution. Correlations in 
the configuration space produce an enhancement of the 
probability that a test particle is subject to a large force, as 
can be seen from Figs.~7 and ~8. 
The disagreement between 
Kandrup's distribution and the experimental one is resolved 
using a theoretical distribution for clustered systems. 
\begin{flushleft}
{\bf Conclusions.}
\end{flushleft}
In this paper the theoretical stochastic force distribution 
in inhomogeneous systems has been studied 
(Kandrup 1980) using a system of 8000 particles 
distributed according to an isotropic Plummer model. The theoretical 
distribution has been compared to that of stochastic force obtained from 
the system evolved over 150 dynamical times. The results of the comparison 
has shown that Kandrup's theory describes correctly the stochastic force in 
inhomogeneous, evolved systems as long as correlations are not present. 
Disagreements between the theoretical distribution and the experimental one, 
observed during the evolution, are due to correlations arising during 
the evolution of the system. The presence of correlations has been 
shown using 
Miller's (1971) test and the two point correlation function has been 
calculated 
using the count in cell analysis (Mo, 1991; Ripley 1980) and it has been 
used in the AB92 theory 
to calculate a theoretical distribution of the stochastic force taking into  
account the correlation effects. The final comparison has shown a good 
agreement between the last theoretical distribution and the experimental 
distribution. Then the stochastic force in inhomogeneous unclustered systems 
are well described by Kandrup's theory while when correlations are present 
it must be substituted by 
the AB92 theory. \\
In conclusion I observe that the role 
of the stochastic force in the evolution of gravitational 
systems increases when correlations are present 
because in such systems there is a greater probability that stars 
are subject to large stochastic forces with respect to the uncorrelated 
one. 
\begin{flushleft}
{\it Acknowledgements.\\} 
\end{flushleft}
I would like to thank my supervisor, V. Antonuccio-Delogu, 
for having suggested the subject of this research and 
for a careful reading of the manuscript.
%


\begin{thebibliography}{}
\bibitem{ahm} Ahmad, A., Cohen, L., 1973, Ap.J. 179, 885
\bibitem{ahm} Ahmad, A., Cohen, L., 1974, Ap.J., 188, 469
\bibitem{ant} Antonov, V. A., 1973, In The Dynamics of Galaxies and star clusters, Ed. G.B. Omarov, p. 139. Alma Ata: Nauka[translated in de zeeuw (1987)]
\bibitem{ant} Antonuccio-Delogu, V., Atrio-Barandela, F., 1992, Ap.J. 392, 403
\bibitem{bar} Barnes, J., 1986, in Dynamics of Star Clusters, Iau Simposium N. 113, ed. J. Goodman \& P. Hut, p. 297. Dordrecht , Reidel
\bibitem{cha} Chandrasekhar, S., von Neumann, J., 1942, Ap.J. 95, 489
\bibitem{cha} Chandrasekhar, S., 1943, Ap.J. 97, 255
\bibitem{del} Del Popolo, A., 1995, A \& A in press
\bibitem{dor} Doremus, J.P., Feix, M.R., Baumann, G., 1971, Phys. Rev. Lett., 26, 725
\bibitem{edd} Eddington, A.S., 1916, M.N.R.A.S., 76, 572
\bibitem{gil} Gilbert, I., 1970, Ap.J., 159, 239
\bibitem{hen} Henon, M., 1973, A \& A, 24, 229
\bibitem{her} Hernquist, L., 1987, Ap.J. Supp. Ser. 64, 715 
\bibitem{hun} Hunger, K., Larenz, R.W., Wilke, K., 1965 Proceedings Second Harvard-Smithsonian Conference on stellar Atmospheres, Cambridge, Mass.
\bibitem{kan} Kandrup, H.E., 1980, Phys. Rep. 63, n.1 1
\bibitem{mer} Merrit, D., Aguilar, L.A., 1985, M.N.R.A.S. 217, 787 
\bibitem{mer} Merrit, D., 1985, Astron. J., 90 (6), 1027  
\bibitem{mil} Miller, R.H., 1971, Ap.J. 165, 391
\bibitem{mo} Mo, H., 1991, Ph.D Thesis.
\bibitem{plu} Plummer H.C., 1911, M.N.R.A.S. 71, 460
\bibitem{pri} Prigogine, I., Severne, G., 1966, Physica, 32, 1234
\bibitem{ryp} Ripley, B.D., 1981, Spatial Statistics, Wiley
\end{thebibliography}
\end{document}